\documentclass[conference]{IEEEtran}
\IEEEoverridecommandlockouts
\usepackage{cite}

\usepackage[dvipsnames]{xcolor}
\usepackage{amsmath,amssymb,amsfonts}
\usepackage{algorithmic}
\usepackage{graphicx}
\usepackage{textcomp}
\usepackage{xcolor}
\usepackage[]{algorithm2e}
\usepackage{caption}
\usepackage{enumerate}
\def\BibTeX{{\rm B\kern-.05em{\sc i\kern-.025em b}\kern-.08em
    T\kern-.1667em\lower.7ex\hbox{E}\kern-.125emX}}
\begin{document}

\title{A Stochastic Graph-based Model for the Simulation of SARS-CoV-2 Transmission}

\DeclareRobustCommand*{\IEEEauthorrefmark}[1]{%
  \raisebox{0pt}[0pt][0pt]{\textsuperscript{\footnotesize #1}}%
}

\author{\IEEEauthorblockN{Christos~Chondros\IEEEauthorrefmark{}, Stavros~D.~Nikolopoulos\IEEEauthorrefmark{*} and
Iosif~Polenakis\IEEEauthorrefmark{}}
\IEEEauthorblockA{\it Department of Computer Science and Engineering \\
 University of Ioannina\\
Ioannina, Greece\\\\
Corresponding Author: \IEEEauthorrefmark{}stavros@cs.uoi.gr}}

\maketitle

\begin{abstract}
In this work we propose the design principles of a stochastic graph-based model for the simulation of SARS-CoV-2 transmission. The proposed approach incorporates three sub-models, namely, the spatial model, the mobility model, and the propagation model, in order to develop a realistic environment for the study of the properties exhibited by the spread of SARS-CoV-2. The spatial model converts images of real cities taken from Google Maps into undirected weighted graphs that capture the spatial arrangement of the streets utilized next for the mobility of individuals. The mobility model implements a stochastic agent-based approach, developed in order to assign specific routes to individuals moving in the city, through the use of stochastic processes, utilizing the weights of the underlying graph to deploy shortest path algorithms. The propagation model implements both the epidemiological model and the physical substance of the transmission of an airborne virus considering the transmission parameters of SARS-CoV-2. Finally, we integrate these sub-models in order to derive an integrated framework for the study of the epidemic dynamics exhibited through the transmission of SARS-CoV-2.
\end{abstract}


\section{Introduction}
The recent outbreak of a disease caused by an airborne transmitted pathogen brought to forefront the vital need for more extensive research over the aspect of technologies that could efficiently and effectively provide solutions to such needs. In this work, we studied the parameters that contribute to the spread of such a pathogen among individuals of a population and proposed a stochastic graph-based model for the simulation of the SARS-CoV-2 spread in order to study further insights exhibited throughout the epidemic dynamics of SARS-CoV-2.

\subsection{Preventing Pandemics}
\label{Preventing Pandemics}

In 2019 a new positive-strand RNA virus appears, as a strain of Coronaviridae emerging as a mutation of the known SARS virus, namely the SARS-CoV-2 that caused the so-called COVID-19 disease that finally developed to a pandemic. As described in \cite{ref33} several counter measures there should be adopted as soon as possible to avoid the transmission of the new coronavirus.  Preventing the spread of a disease in order not to be developed in a level such that the vast majority of the population to get infected, i.e. a pandemic occurs, several counter measures have to be taken. Manika and Golden in \cite{ref29} investigating the influenza pandemic caused by H1N1 conclude that the prevention of a pandemic could be more effective than responding to it by mitigating its effects.  

Several pathogens have been transmitted from animals to humans by several pathways \cite{ref30} and by mutating themselves spread among individuals evolved to pandemics. As described in \cite{ref28} the interface between humans and animals is of major importance in the process, suggesting hence a key point in the emerging of pandemics as their investigation may results to the development of prevention strategies. In \cite{ref31} Morse et. al. throughout their investigation over the emerging pathogens regarding their origination, the interface of transmission to humans, etc. conclude that the deployment of mathematical modeling and information technologies can provide a crucial effort in the recognition of new pathogens and fatherly provide new risk assessments. Jung et. al. in \cite{ref32} investigate the case study of influenza pandemic and the role of time-dependent prevention strategies, referencing the major importance of quarantine as a countermeasure for pandemic prevention. 

\subsection{Our Approach}
\label{Our Approach}

Through the recent outbreak of SARS-Cov-2 pandemic COVID-19 several studies and research have been oriented on the defense line against pandemic mitigation.  Encountering several infection cases, from which a significant amount results to ICU and a percentage of them do not survive, our research approach consists another effort to deploy computer science over the sector of public health, and to a further extent to mitigate the viral spread that could be potentially evolved to a pandemic.

In this work, we study the propagation of airborne viruses and provide a method to model their spread in order to deploy graph-based methods preventing the extension of the spread to a pandemic. The stochastic model developed for the study of the spread of airborne viruses, (w.r.t. the dynamics of SARS-CoV-2 spread), deploys three sub-models, namely the spatial model, the mobility model, and the propagation model, responsible for the simulation of the area the population is moving and transmits the virus, the simulation of the mobility patterns followed by the members of a population inside a specific area, and the simulation of the dynamics and the context under which the simulated airborne virus is transmitted among the population, respectively.

\subsection{Contribution}
\label{Contribution}
Throughout this work there has been conducted a research about the epidemiological status of the airborne virus spread, regarding the dynamics of a potential pandemic. The main contribution of this work is the development of a hybrid simulation model representing a complex system behavior. In particular, the proposed simulation framework incorporates a spatial model to simulate the area of a given city utilizing realistic images taken from Google maps, the mobility model that actually is an agent-based model that simulates the interactions of individuals within a population (in our cases individuals that are moving inside the city area), and the propagation model that is a stochastic model that simulates the development of the viral load inside a host and to a further extent it is responsible for the simulation of the propagation of the virus. Furthermore, the proposed framework is a able to incorporate, by embedding various external algorithms, e.g. for the enhancement of contact tracing procedures \cite{refTR}, the implementation of several counter-measures in order to study their effect on the behavior of a pandemic. 

\section{Theoretical Background}
\label{Theoretical Background}

Next, we present the basic theoretical background behind the development of our proposed  simulation framework for pandemic prevention, regarding the basic principles, the underlying modeling for the simulation of the spatial parameters, the mobility models utilized for the stochastic development of mobility patterns and the epidemic models deployed to simulate the propagation of the airborne virus, in our case under the context of COVID-19 pandemic. 

\subsection{Spatial Parameters and Community Transportation}
\label{Spatial Parameters and Community Transportation}

One of the factors of major importance that affects the spread of a contagious disease are the spatial parameters. The congestion of individuals and the corresponding increase on the density of a sample, would definitely accelerate the propagation of a pathogen, and even worst in case of an airborne virus, making this aspect quite motivational for our research and further for the integration of such an approach to our proposed model. 

Through our research, we adopted and extended the approach presented in \cite{ref45} deploying the simulation of the underlying texture that represents the area within witch the individuals are moving, by utilizing real images of cities from Google Maps and, after several image processing procedures, to transform them into graph objects handled by our model to define routes and paths following the streets and the topology of the town planning of the city under consideration.

Through the literature there have been proposed several approaches for the utilization of Agent-based Simulations \cite{ref40, ref41, ref42, ref44}. Among the utilization of stochastic approaches, the Agent-based models meet the requirements for their utilization on the modeling and simulation of community transportation.  Agent-based models provide the ability to deploy underlying stochastic procedures, in order to support decision making and define the interactions among individuals. Such approaches provide the potentials of a more realistic simulation since, as described in \cite{ref5}, they are indicated for the modeling of social networks and special movements and obviously of major importance for the outcome of the pandemic with respect to its basis on the modeled human behavior. 

In our approach, for the specification of our proposed framework for the simulation of the spread of an airborne virus among individuals that move inside the boundaries a of a city area, we developed such a stochastic model that takes into account a set of probabilities that define many aspects of the mobility of the individuals inside the city area, ranging from the crucial factor of how many times daily it is defined for an individual to perform a route inside the city, to the decision of its target location and where or not the individual should contain his route reaching another point of the city or to return to the location defined as its base, as we will discuss more extensively later. 

Finally, another novel point of our proposed model is that in accordance to the proposed mobility model, we also develop a contact graph that represents the potential friendship relation among individuals and stochastically allow any pair of individuals linked with a friendship relation in the corresponding graph to remain for a specific period at close range once they are met during their paths/routes. We are estimating that such an approach, and obviously its combination with our proposed mobility model should definitely reduce the overall distance between our proposed model and the reality.

\subsection{Spread of Airborne Diseases}
\label{Spread of Airborne Diseases}
In order to model the spread of a disease, and more precisely to simulate its propagation among susceptible individuals, an underlying epidemic model is required for the depiction of the states that an individual may be, and additionally the underlying properties that may be required in order to model the physical substance of the transmission of an airborne virus. Next, we discuss both approaches and to a further extent provide an introductory discussion on what we utilize in our proposed approach for the propagation model that is incorporated by our proposed model.

In order to formally simulate the propagation of a spreading disease, there has already been proposed since 1927 by Kermack and McKendrick, the utilization of the so-called compartmental models.  Through this modeling, the states of the parts of the population among witch a disease is spreading, have a specific labeling, namely Susceptible, Infected and Recovered, consisting the SIR epidemic model, that describes the transition over whose states (i.e., S, I, R) an individual ma be during the spreading of a disease. That is, modeling the spread of a disease that any individual is Susceptible to the pathogen, then after his infection by the pathogen is labeled as Infected and finally after his treatment or recover turns to the state Recovered, is preferable to model disease that once a Susceptible individual gets the disease, turning for a specific period to the state of  Infected, after his treatment the individual is Recovered, been no longer in the Susceptible state, without however the exclusion of being Immune to the disease, that is described by another epidemic model. 

The SIR epidemic model has several extensions that are adapted to describe more accurately diseases that exhibit a more specific behavior. For example, if an epidemic model is required for the description of state transition of influenza, then knowing that an individual may get Infected more than once, the SIS epidemic model may be deployed to describe the behavior of such a disease \cite{ref35, ref35b}. In that model the third state corresponds to the state of Susceptible (i.e., Susceptible, Infected, Susceptible) in order to permit the transition from the state of Infected to the state of Susceptible allowing the probability of getting Infected again. 

In case of the existence of an incubation period, an extension of the SIR epidemic model, the SEIR epidemic model may be applied to describe the transitions of the population among these states (i.e., Susceptible, Exposed, Infected, Recovered), including also the state Exposed \cite{ref36}, in order to describe the transitional state between the first date of infection and the day that symptoms or any other metrics shows that the individual in Infected.

Regarding the specifications exhibited by the spread of the SARS-CoV-2 coronavirus we deploy the SIII epidemic model, including the states (Susceptible, Infected, Infectious, and Immune), where an individual primarily is Susceptible to get Infected by the virus, after the exposition to the virus the individual is getting Infected and from a short time interval (i.e., 4-5 days from exposure to symptoms onset \cite{ref37,ref38,ref39}) and after starts to be contagious, and hence being Infectious, while after a period of days he recovers being Immune and not contagious. Since there is not enough information from research in the literature about reported cases of reinfection, at least from the same strain of the SARS-CoV-2, we estimate that the SIII model fits perfectly meeting the requirements of our proposed model.

Finally, concerning the physical substance of the transmission of airborne viruses in our approach we developed and propose a model that takes into account the main pathways over which the new coronavirus is transmitted, which is the infected droplets. Throughout this aspect we define two main procedures, namely, the evolution of the viral load inside the infected hosts that consequently defines the percentage of the infected droplets emitted by an infectious host during actions like coughing, sneezing, talking, or even breathing and the spatial method that these droplets are transmitted to susceptible individuals.  Hence, having modeled and, as close as possible to reality simulated, the evolution of the viral load, and to a further extent the percentage of the infected droplets emitted, we compute stochastically a probability for an infected individual to perform an action, e.g., coughing, considering the corresponding volume of emitted droplets. Then, we compute the number of susceptible individuals that are located proximate to this infectious individual and computing a geometrical representation of the produced cloud of droplets in a specific distance we stochastically infect some of the individuals.

\section{The Proposed Model}
\label{The Proposed Model}
In this section we discuss the basic components of our model and present more extensively the incorporated models, namely, the spatial model, the mobility model and the propagation model. 

\subsection{Model Overview}
\label{Model Overview}
The proposed model for the simulation of the spread of an airborne pathogen among individuals that are moving inside a specific area of a city, is consisted namely by the spatial model, the mobility model, and the propagation model. The spatial model simulates the structure of a given city regarding its topology with respect to the streets and the corresponding town's planning.  The mobility model acts as an agent-based modeling approach for the simulation of the mobility patterns (i.e., the routes) followed by the individuals (in our case the citizens of a town) in order to move between specific points of the city, Finally, the propagation model, is responsible for the transmission of the airborne pathogen and to a further extent for the deployment of the underlying epidemic model alongside the airborne pathogen transmission specifications, in our case the ones exhibited by SARS-CoV-2 virus.

\subsection{Spatial Model}
\label{Spatial Model} 
Next, we present the architecture of the proposed spatial model, discussing the transformation of a Google-Map image to an object that depicts the structure of a city, modeling among others the orientations exhibited by the city roads, the points of interest that attract the individuals' routes, and next the construction of the underlying undirected weighted graph utilized for the construction of the paths in the corresponding routes. \\

\subsubsection{\textbf{From Google-maps Images to Graphs}}
\label{Transforming G-map Images to Undirected Graphs}

\begin{figure*}[t!]
\centering
\resizebox*{4.24 cm}{!}{\includegraphics{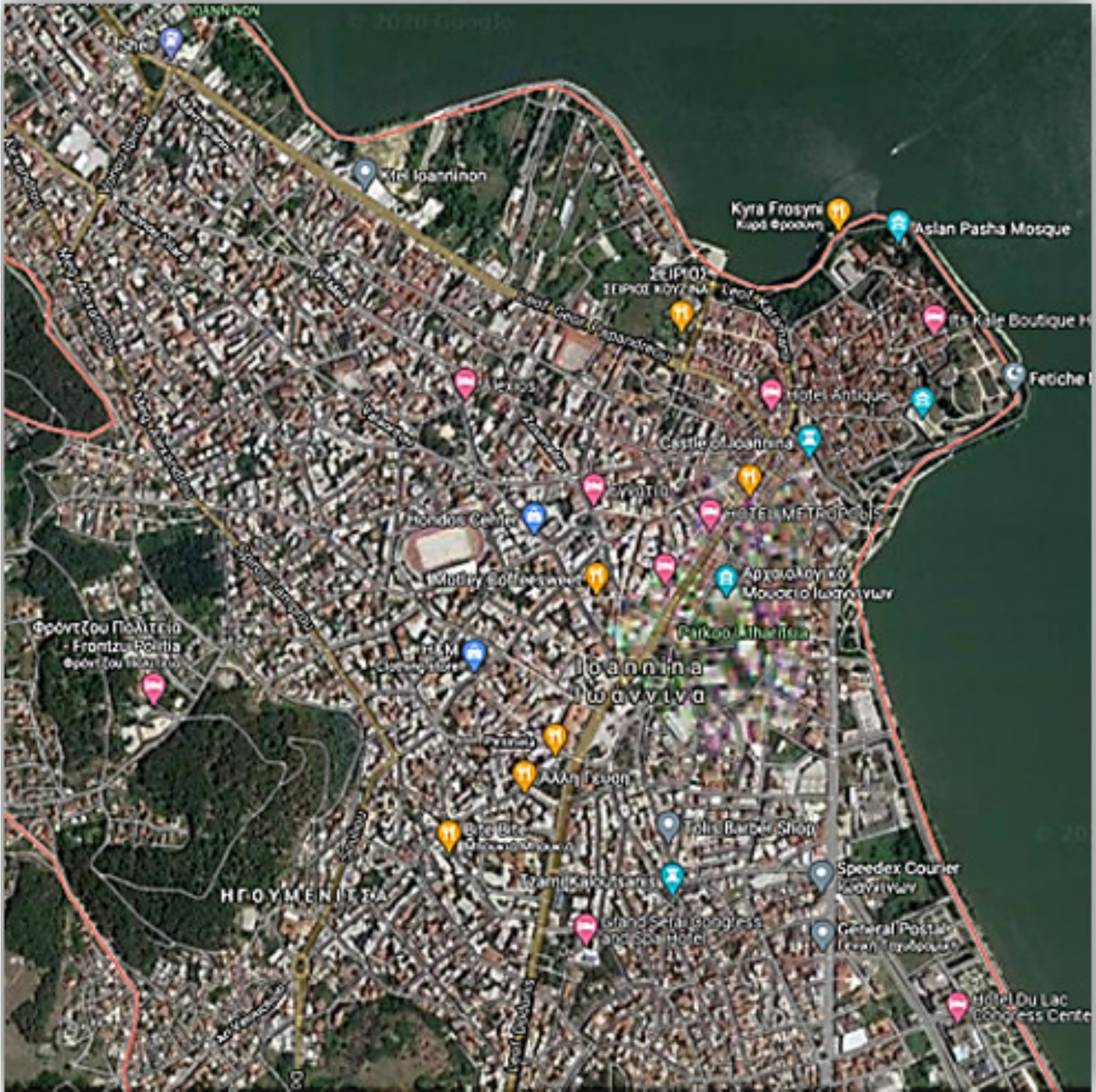}}
\resizebox*{4.2 cm}{!}{\includegraphics{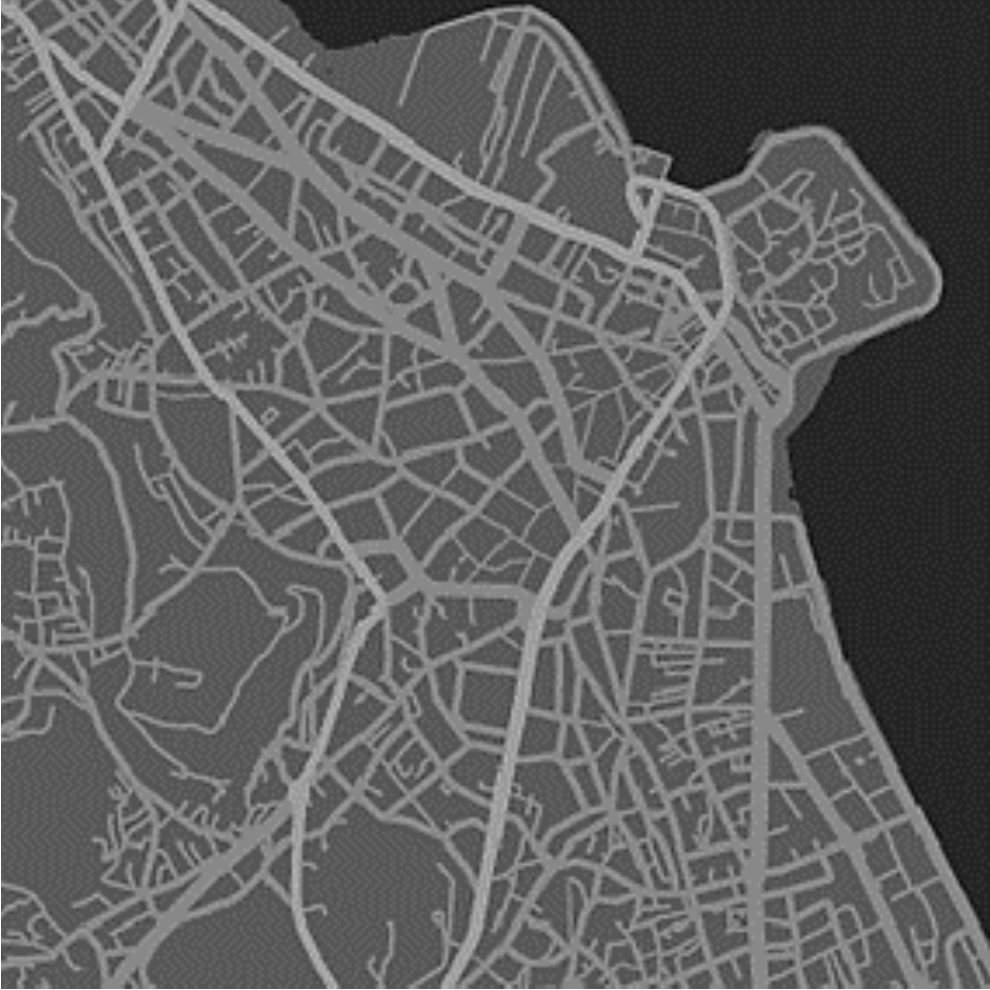}}
\resizebox*{4.16 cm}{!}{\includegraphics{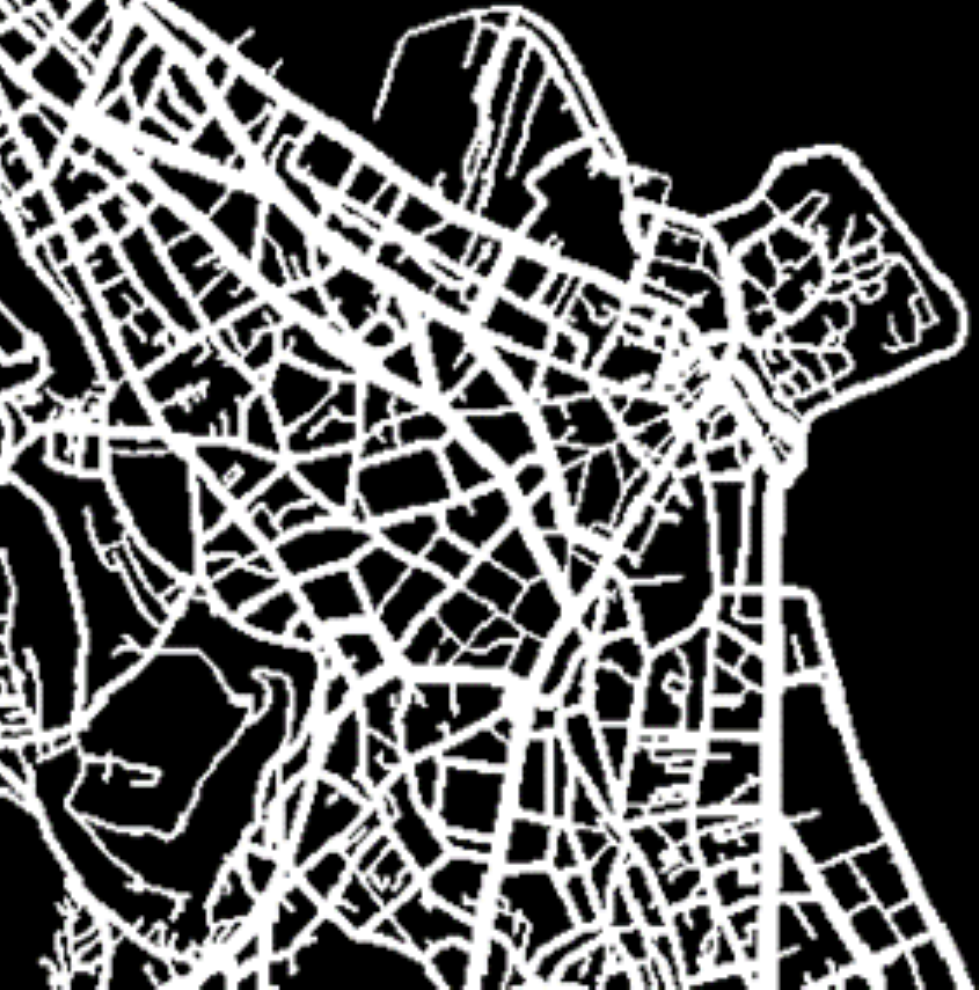}}
\resizebox*{4.38 cm}{!}{\includegraphics{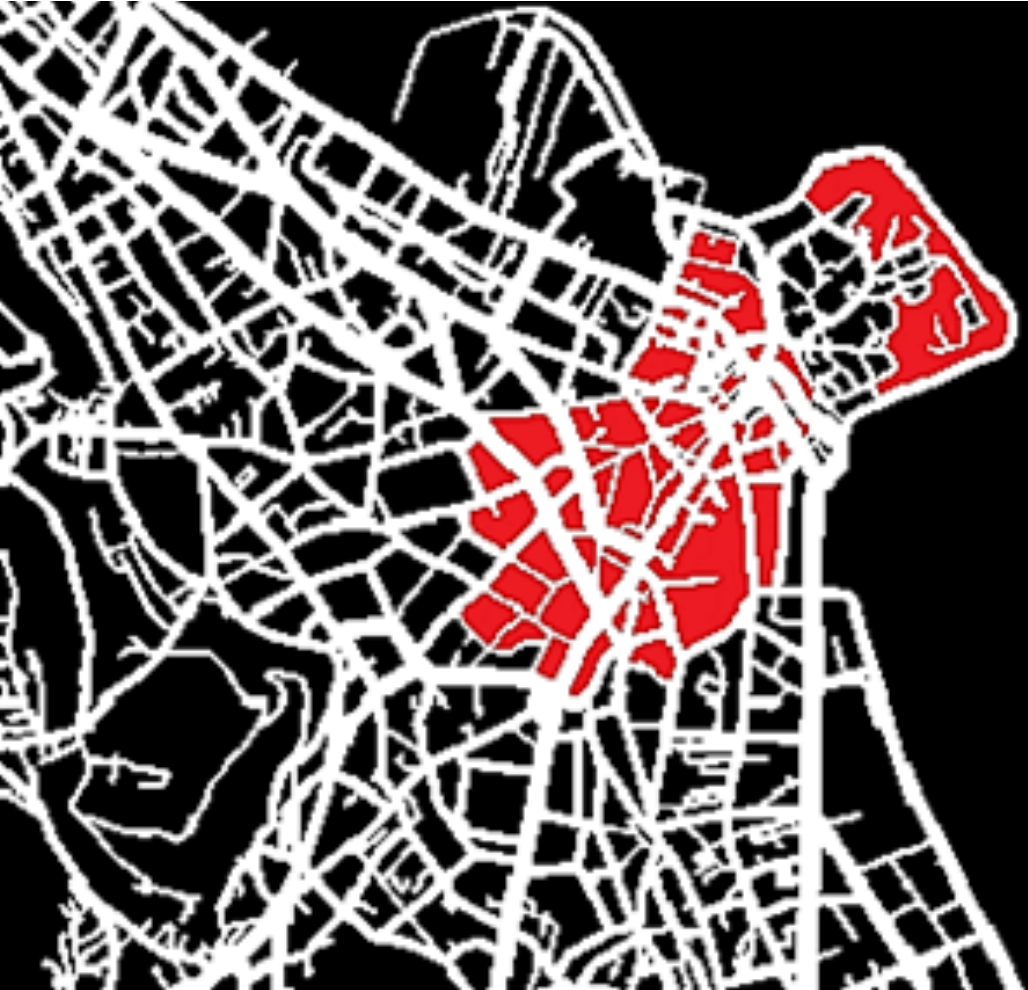}}
\hspace{0.2 in}
\phantom{}(a) Google Maps Image.  \phantom{xxxx}(b) Gray-scale Image. \phantom{xxxxxxxxx}(c) B/W Image. \phantom{xxxxxxxxxx}(d) PoI Assignment.
\caption{Example of the image transformation procedure.} \label{image-trnsf}
\end{figure*}

Through our research, we adopted  and extended the approach presented in \cite{ref45} deploying the simulation of the underlying texture that represents the area within witch the individuals are moving, by utilizing real images of cities from Google Maps and, after several image processing procedures, to transform them into graph objects handled by our model to define routes and paths following the streets and the topology of the town planning of the city under consideration. The purpose of this operation is a detailed graph-based representation of the image’s roads, that model the roads of the city, required for the simulation. At the end of the transformation, an undirected graph is constructed, whose vertex set correspond to a square spot of the street (i.e., pixel of the image), while an edge belongs to the edge set of the graph if the corresponding end points are neighboring square spots (i.e., $7-$neighborhood) of the street.

The proposed model transforms any given image of any size into a graph alongside all the auxiliary information required. Our main goal is to create a graph whose vertex set corresponds to the streets from the given Google Map image, modeling the area in which the propagation is investigated. More precisely, we select the preferred area from Google Maps see Figure~\ref{image-trnsf}(a), capturing the corresponding segment cropping it as an image, and finally omitting the irrelevant information, i.e., tags, names etc.

Next, we transform the image into a $2-$D matrix, with each particular cell containing three different values (i.e., the RGB color index of each pixel). In order to distinguish the cells that correspond to streets and the cells that correspond to buildings, we first transform the initial image into a grey-scale image, casting the RGB index value of each cell into a single value that represents the intensity of gray color, and applying a thresholding procedure defined by specific image characteristics, see Figure~\ref{image-trnsf}(b). 

In our approach we consider that the cell values of the matrix above the specified threshold value refer to square spots of streets, while cell values of the matrix below the specified threshold value refer to square spots of the area outside the streets (i.e., buildings, obstacles, etc.) representing non-accessible points. The application of this procedure results to a $2$-D matrix full of $0$ and $255$ and a corresponding black-white image that represents the initial image with streets colored with white color and buildings colored with black color, see Figure~\ref{image-trnsf}(c). Considering the two-dimensional matrix full of $0$ and $255$ the actual graph-construction procedure begins by transforming each particular cell whose value is $255$ (i.e., a cell that represents a square spot of the street) into a vertex of the graph. Alongside the creation of the graph vertices, the same procedure constructs the edges of the graph considering the neighbors of the cells, taking into account the $7$-neighborhood, that also have value equal to $255$. The result is a detailed graph-based representation of the initial image, denoted as $G_{map}$, whose vertices represent the actual square spots of the streets, while its edges represent the transitions (i.e., steps) among them. \\

\subsubsection{\textbf{Assign Points of Interests}}
\label{Assign Points of Interests}

To make our model more realistic, we consider factors related to the high traffic and the mobility of any particular city. Points of interests or attraction points, or for short, PoI, in a city are frequented or high-level traffic areas which increase the congestion and hence the hyper-transmission of the airborne pathogens. In our model, the user assigns the points of interest (or, for short, PoI) at the start of the experiment, by marking those areas in the initial image representing the high traffic areas of the city, see Figure~\ref{image-trnsf}(d). To this point, it is of major importance to notice that only the non-accessible points of the image (i.e., buildings and obstacles) need to be marked in order to assign the importance level that denotes the points of interest. At the end of this procedure, to simulate the traffic of the initial city’s roads, our proposed model constructs a new matrix, the so-called Attraction-Matrix, whose cells correspond to square spots of the streets with a value from a specific range of numbers (from $0$ to $10$), where the higher the value of the cell, the more frequent the square spot of the street is. 

\noindent In our model, distinguish the street spots as follows:
\begin{itemize}
\item a {\tt hot spot}, a square spot of the street or a particular cell of the Attraction-Matrix, whose value in the corresponding cell ranges between $10$ and $7$, i.e., $(7,10]$. 
\item a {\tt warm spot}, a square spot of the street or a particular cell of the Attraction-Matrix, a cell whose the only if the corresponding cell has value that ranges between $7$ and $4$, i.e., $[4,7]$, and 
\item a {\tt cold spot}, a square spot or a particular cell of the Attraction-Matrix if and only if the value of the corresponding cell between $3$ and $0$, i.e., $[0,4)$. \\
\end{itemize}

\begin{figure}[t!]
\centering
\includegraphics[scale=0.5]{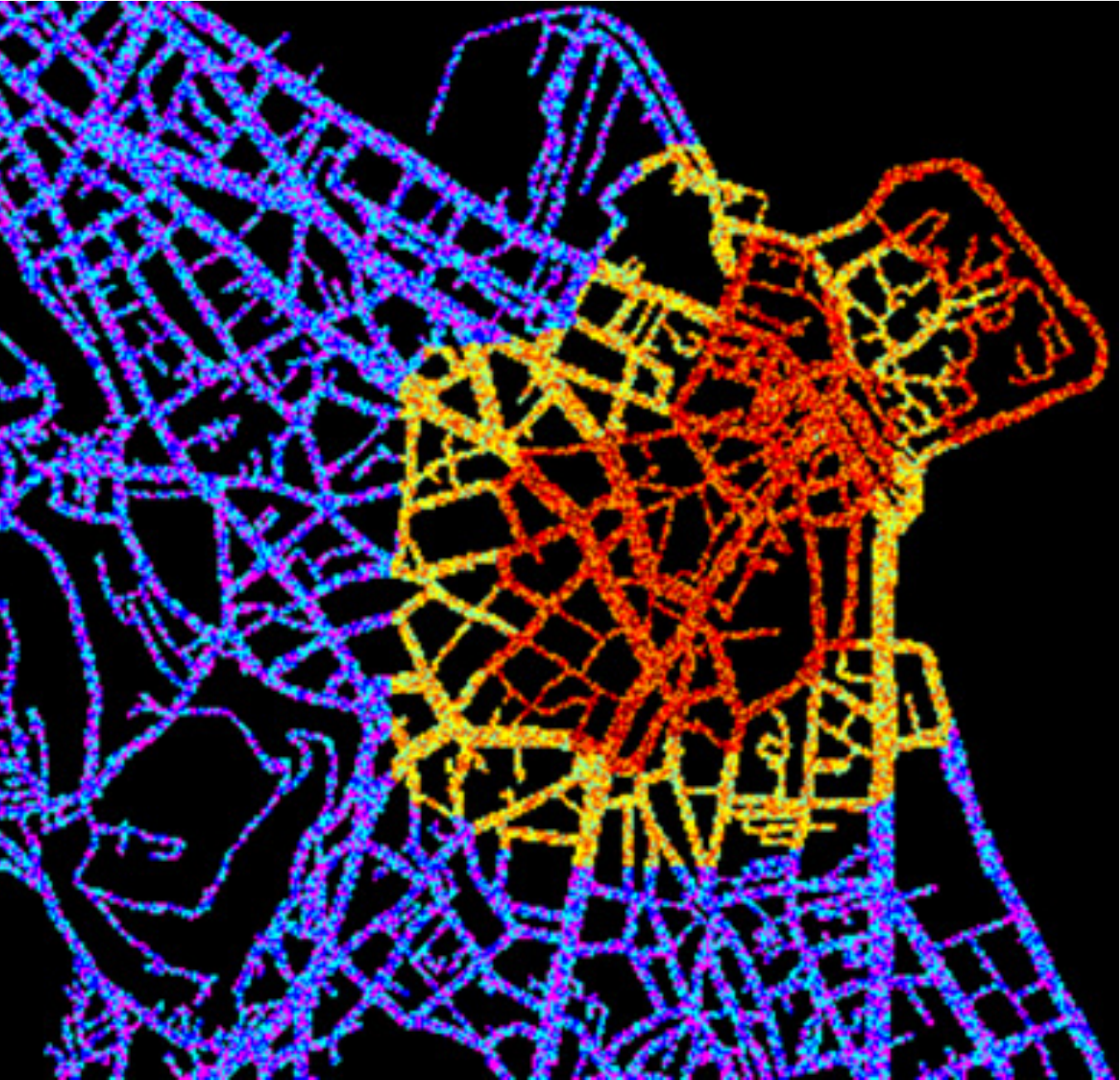}
\caption{Assignment of Hot, Warm, and Cold Spots.} \label{routes1}
\end{figure}

Once the PoI have been marked to the black and white image, we compute the significance of each particular cell which corresponds to a square spot of the street of the Attraction-Matrix considering its distance from its closest PoI. The cells will be defined as hot, warm, or cold spots according to their distance from their closest PoI. To this end, it is worth noting the Attraction-Matrix is the major component in the construction of the route basis and the determination of which path each individual follows, moving from a source to a destination point, see Figure~\ref{routes1}. \\

\subsubsection{\textbf{Constructing the Route Basis}}
\label{Constructing the Route Basis}
In our model, each road is a set of pixels of the initial image, that have been transformed to the corresponding cells of the Attraction matrix which holds a value that ranges between 0 and 10 (i.e., [0,4) for cold spots, [4,7] for warm spots and (7,10] for hot spot). Considering the undirected graph $G_{map}$, each vertex represents a square spot of the street and corresponds to a cell in the Attraction-Matrix. At this point, in order to construct the route basis, we need to reconstruct the weights of the underlying undirected graph that so far have weights equal to $1$, considering the PoI that have been set, recalling that its vertex set corresponds to street spots and its edge set corresponds to their in between neighbors (7-neighborhood).

 The new weight for an edge $e_{i,j}$ connecting the vertices $v_i$ and $v_j$, such that $e_{i,j} \in E(G_{map})$ and $v_i, v_j \in V(G_{map})$ is calculated as: 

\begin{equation}\label{eq1}
w(e_(i,j))=10 – L_i+10–L_j+1,
\end{equation}

\noindent where $L_i$ and $L_j$ are the Attraction-Matrix cell’s values (i.e., [0-10]), based on the vertices connected by the corresponding edge. 

For example, let that we want to calculate the weight of the edge, $e_1$. The procedure begins by finding the IDs of the nodes that $e_1$ connects. In our example let us assume that the vertices with $ID_{L_j}=(324,51)$ and $ID_{L_i}=(323,50)$ and are connected with the edge $e_1$. The procedure continues by searching the values in the corresponding cells of the Attraction Matrix (i.e., in this example $L_i$ is the value of the cell with coordinates $i=323$ and $j=50$ and $L_j$ is the value of the cell with coordinates $i=324$ and $j=51$). 

\begin{figure}[t!]
\centering
\includegraphics[scale=0.5]{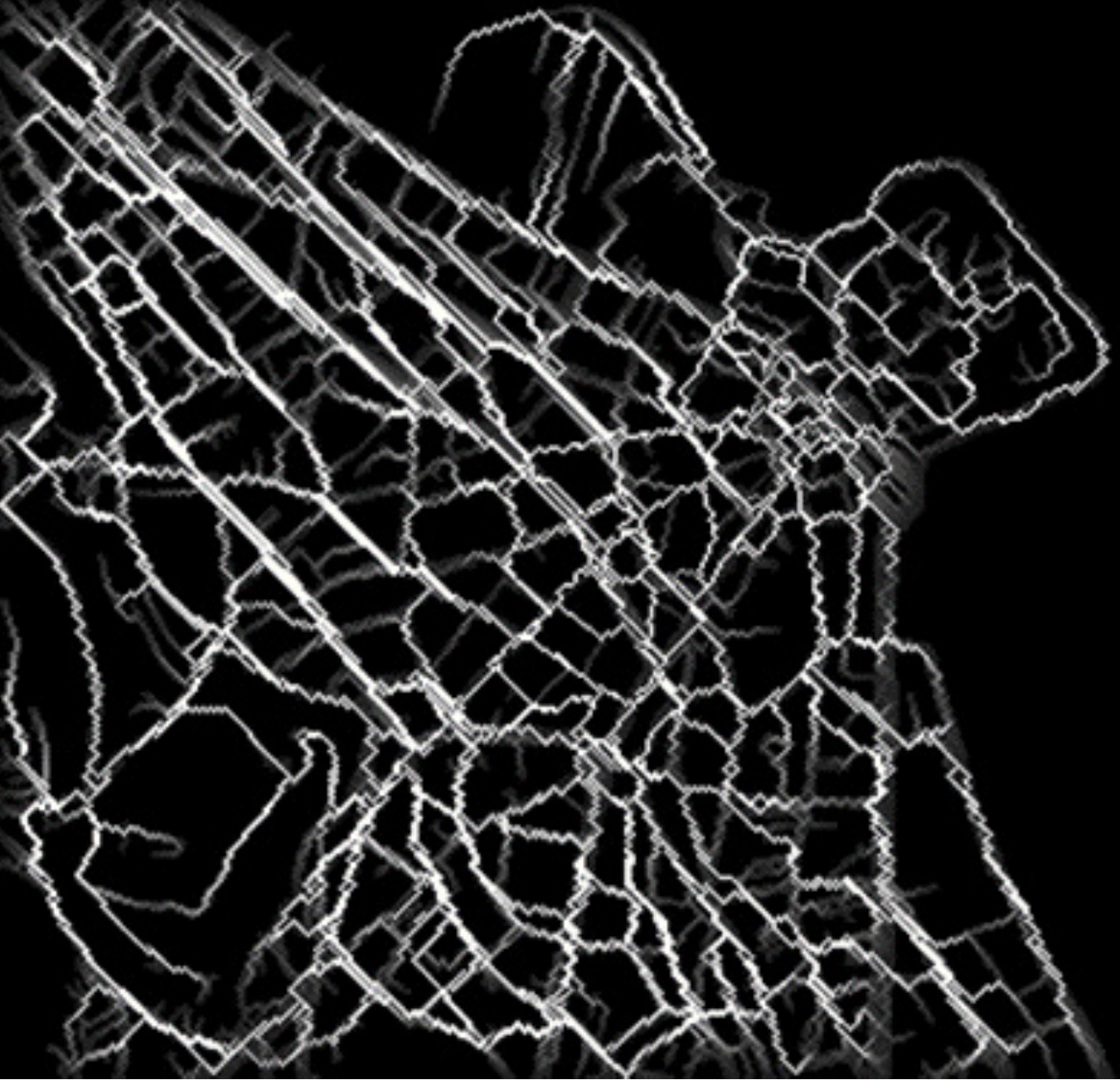}
\caption{Frequently walked streets from the route-patterns created by individuals.} \label{routes2}
\end{figure}

To this point, it is important to note that in the end of the procedure all the weights of the graph will be recalculated, and according to our formula their corresponding values will be set in the range $[1,21]$. Considering two hot spots whose values are 10, using our formula to calculate the weight of the corresponding edge the output will be 1. Additionally, considering two cold spots whose values are 0 our formula will calculate the weight of the corresponding edge as 21. Hence, the proposed model reconstructs the edges of the undirected graph considering the PoI of the image based in in Equation~\ref{eq1} in order to construct the basis for the possible routes, (see Figure~\ref{routes2}).

\subsection{Mobility Model}
\label{Mobility Model}

Our proposed mobility model is developed to simulate the movements in terms of mobility of the individuals, the frequency of their walks and the paths they select to follow to reach their destinations, as to be major factors that affect the dynamic density of the city’s network considering the congestion of the individuals in segments of the city area, and to a further extent to the dynamics of the spread of on airborne transmitted pathogen. Next,we describe the scheduling of the individual’s mobility and the algorithms used to compute the paths followed by the individuals.

\begin{figure}[t!]
\centering
\includegraphics[scale=0.6]{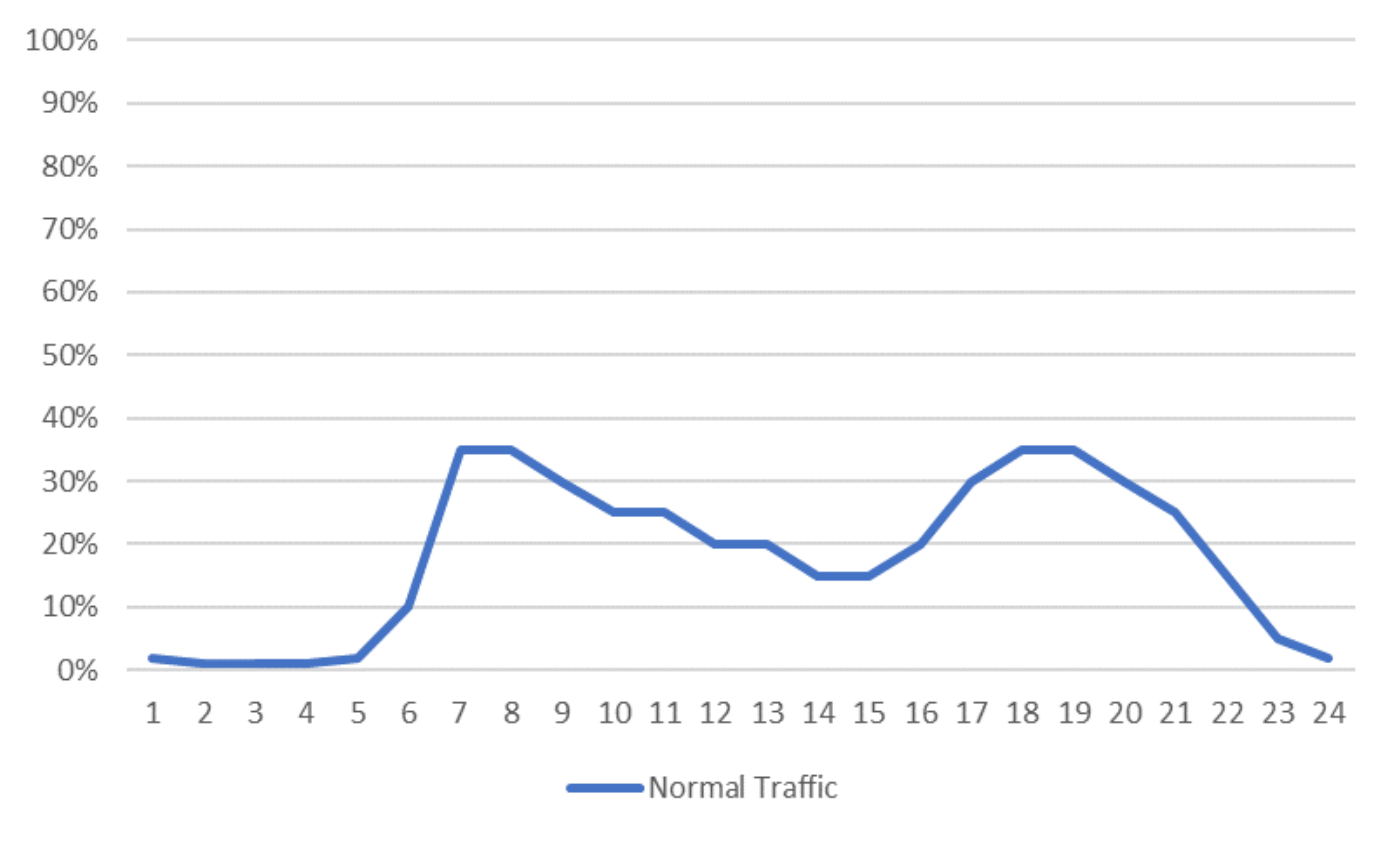}
\caption{Hourly congestion in population mobility inside the boundaries of a city.} \label{rush}
\end{figure}

Considering that the mobility of the population inside a city area is not totally random, and taking into account that rush hours may differ across different cities, the proposed model is flexible enough to support variations exhibited in any hour peaks, according to simulation demands, w.r.t. the city under consideration. More precisely, every single individual has a probability to start a route at any time during the day from a source point of the city to a random destination point. The probability for each individual to start moving into the city increases or decreases according to the percentage of the population estimated regarding specific time-periods in a day (i.e., rush hour or not). More precisely, based on the data depicted in Figure~\ref{rush} where the $x-$axis represents the time-period of a day in hourly basis, and the $y-$axis represents the percentage of the population moving inside a city area. Note that, in our case, considering the curve depicted in Figure~\ref{rush}, we utilized data collected by the national transportation organization.

Next, we discuss the procedure that describes the assignment of the source and the destination points defining the paths to be followed by each individual (i.e., the routes that model the mobility of the individuals inside a city area), how points of interests affect those paths and possible alternative route assignments as long as a path has already been set. 

As we can see in Figure~\ref{rush} that depict the hourly transportation traffic for a specific city, obtained by data provided by the national transportation organization, which represents the percentage of the individuals moving in a city over specific time-periods, our model schedules the mobility of the individuals providing simulation of individual’s movements as close as possible to reality. 

Having constructed the $G_{map}$ from our initial image whose vertices represents square spots of the city routes and weighted edges represent the connections between them. The weight of each edge depicts the characteristics of the corresponding vertices (i.e., a very low weighted edge links two vertices, that possibly correspond to two neighboring hot spots in the initial image). Given a source and a destination point as vertices of the $G_{map}$, the proposed mobility model allows each individual to select the less weighted (i.e., shortest) path among all possible paths that connect the source-destination points. Given a source and a destination points, the shortest, and hence the less weighted, paths in our graph correspond to the frequently walked paths of the city, (see, Figure~\ref{routes2}). 

Additionally, it is of major importance to note that this approach corresponds these paths to the paths that utilize the edges among the spots with greater values, or, equivalently to the paths that contain the edges that are included in most of the paths. To achieve this, we use Dijkstra’s shortest path algorithm for every individual who decides to move from a source to a destination point of the graph, ensuring hence that the selected path includes the less weighted edges. \\

  \begin{table*}[t!]
\renewcommand{\arraystretch}{1.6}
\centering
\caption{Probabilities of actions and corresponding volume of emitted droplets per healthy individual.}
{\begin{tabular}{|l|c|c|c|c|}\hline

\textbf{Actions performed by an individual:} & \textbf{Breath} & \textbf{Cough} & \textbf{Sneeze} & \textbf{Talk} \\ \hline
\textbf{Probability of performing an action}  & 0.8    & 0.1   & 0.05   & 0.05 \\ \hline
\textbf{Volume of Emitted Droplets (VoED)} & 0.2$m^3$    & 2$m^3$   & 3$m^3$    & 0.5$m^3$  \\ \hline
\end{tabular}}
\label{prob-table-a}
\end{table*}
  
\subsubsection{\textbf{Assignment of Source-Destination Points}} 
  The procedure of the assignment of start and destination points in a route for each individual is crucial and of major importance for the transmission of an airborne pathogen. As long as the points of interests have been set and the weighted edges of the graph have been reconstructed, our system assigns to each individual a static source point that remains immutable, representing e.g., the residence of an individual. The source point of an individual is a point in the map that the individual starts and finishes the most of his routes representing the location of his home. Then, iteratively over specific periods of time every individual has a probability to make an action (i.e., to define a path between its source and a destination point and start the corresponding route through the city) depending on his corresponding state (i.e., ``stand", or ``move").  At the beginning of a time-period, all the individuals are located at their residences (i.e., home locations), so each individual computes probabilistically on whether to stay at home or to start a route in the proximal time-period. 
  
  Additionally, in simulation terms, a ``period" refers to the simulation of a day, while the ``time-period" refers to the next hour. Then, each individual whose route is scheduled to start, a random time-slot ranging from 1 to 60 minutes from the current time (in our case, 1-500 simulation steps correspond to one hour of the day) is assigned, in order to start its route.  Moreover, it is worth noting that when the individual is about to move, w.r.t the time defined by the assigned time-slot, a random destination point of the city’s map is assigned in order to construct the route of an individual that is about to move. Since the destination point has been set, we deploy Dijkstra’s shortest path algorithm in order to calculate the shortest path of the graph and navigate the corresponding individual between source and destination points via shortest and less weighted path. This procedure described some of the basic actions of people that our system supports through the mobility model. \\

\subsubsection{\textbf{Optional Selection of Sequential Routes}} 
   As we mentioned previously, our proposed model decides whether a individual starts a route or not, within a proximal time-period w.r.t the probability defined for the corresponding time intervals depicted in Figure~\ref{rush}. Focusing on possible actions that may occur during the route of an individual, when it is time for an individual to start a route, the proposed model calculates the shortest path and the corresponding individual is placed onto the vertex of the graph that represents his source point (i.e., residence of this individual) in the city. At this point, in every simulation step the individual moves from his current vertex of the undirected weighted graph to one of its neighbors, more specifically, to the next vertex that Dijkstra algorithm undermines for the corresponding path. However, during the navigation on the streets, as in reality, there is a probability of meeting another known individual who also follows the route probably to the own destination point into the graph (e.g. routes from pairs of different source-destination pairs that are crossed, or paths with joint edges). In that case, both individuals will stay at the same vertex for $0$ to $5$ time slots and then they will continue their navigation to their corresponding destination points. 
   
   In our model in order to construct such a contact-network we construct a random graph, depicting the friendships among the individuals. This procedure reflects a real-life scenario where two people that know each other randomly meet in a city’s street and stand-by for at least a handshake, or a short talk. As long as the individual reaches the destination point of the city, remains stable to the corresponding vertex of the graph for a specific period of time and then decides one of the following actions, i.e., whether to move to a new destination point of the city, or to set the destination point to the home location, i.e., the new destination point is reset to the initial start point. With this approach, our model is able to simulate real-life cases where during a route, an individual may meet a friendly contact and pause his walk for a short amount of time, considering also the probability that a walk may include multiple destination points before an individual return to source location.

\begin{figure*}[t!]
\centering
\resizebox*{4.4 cm}{!}{\includegraphics{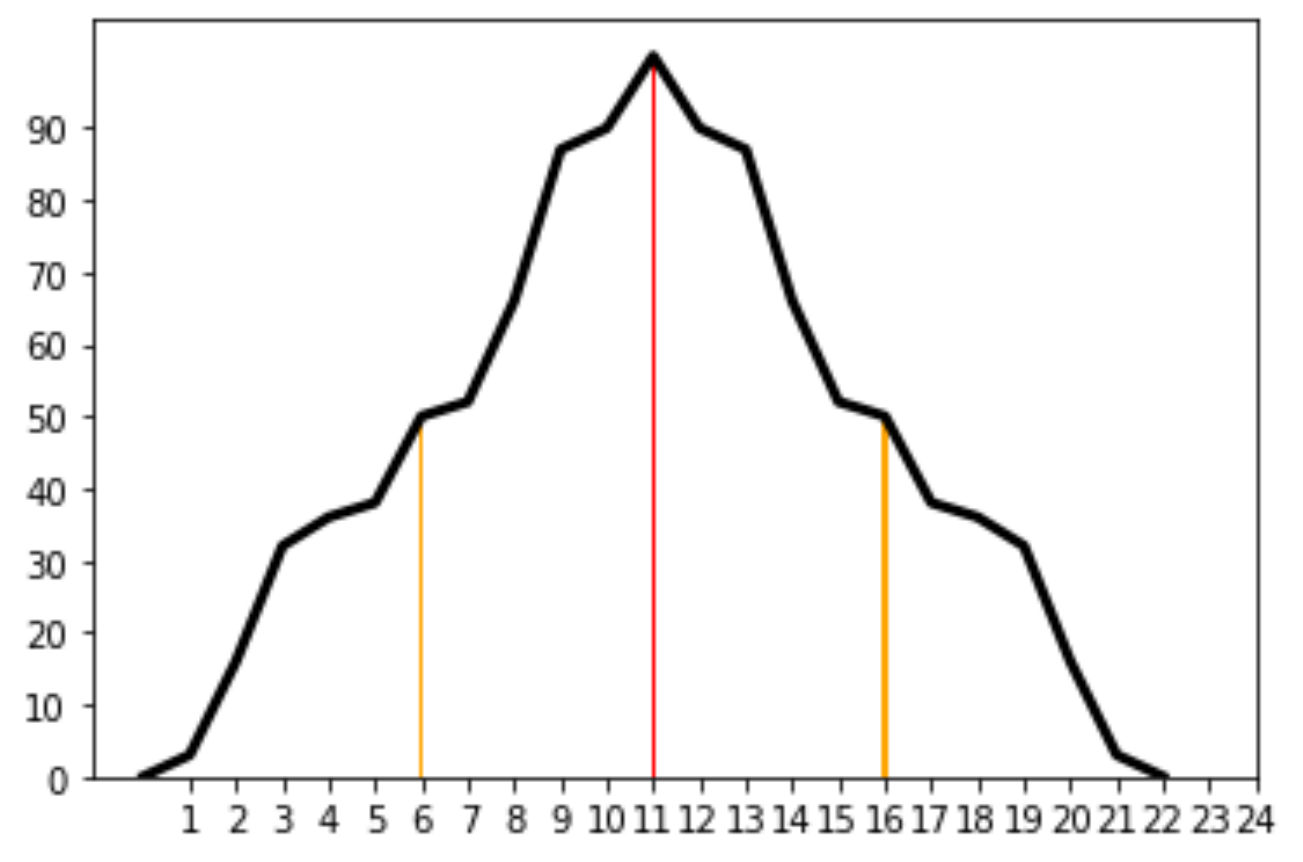}}
\resizebox*{4.4 cm}{!}{\includegraphics{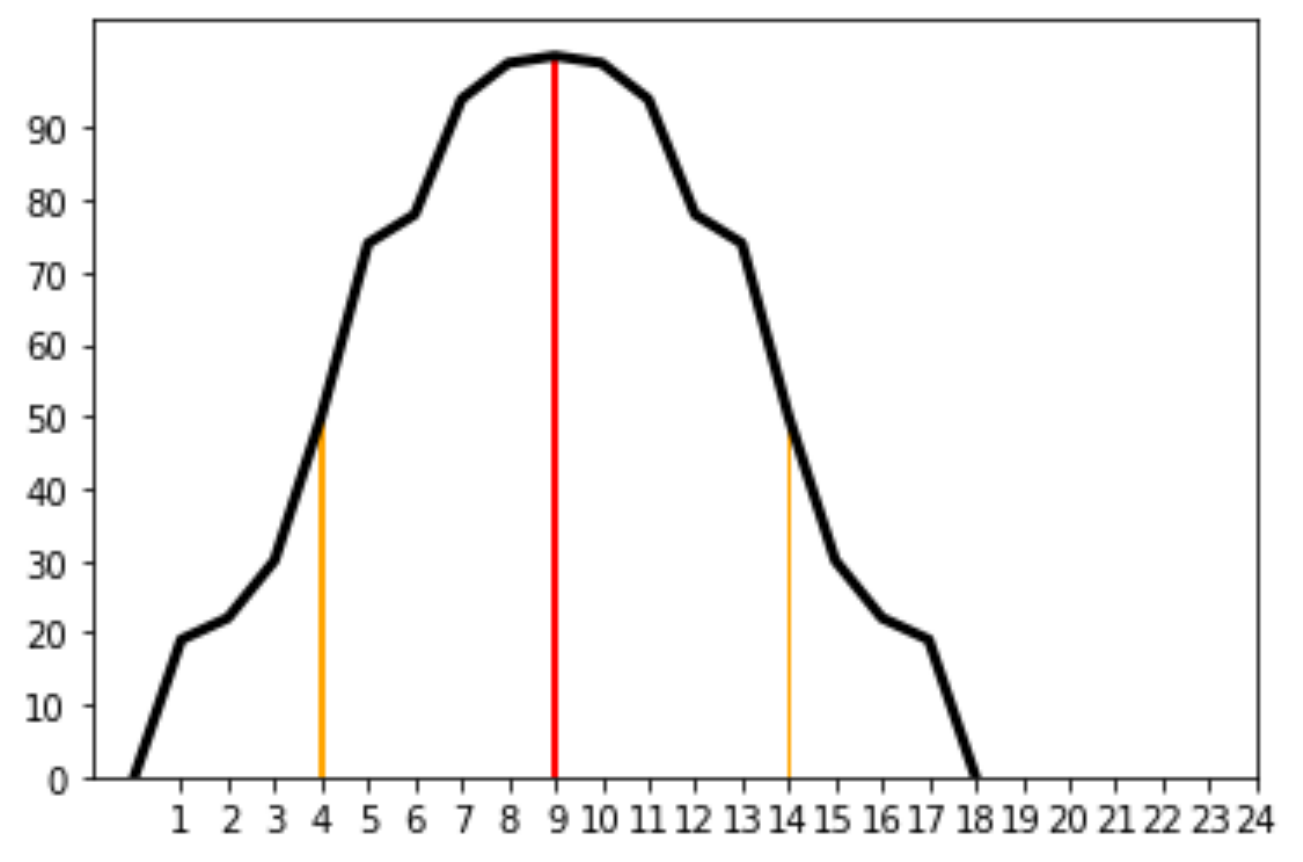}}
\resizebox*{4.4 cm}{!}{\includegraphics{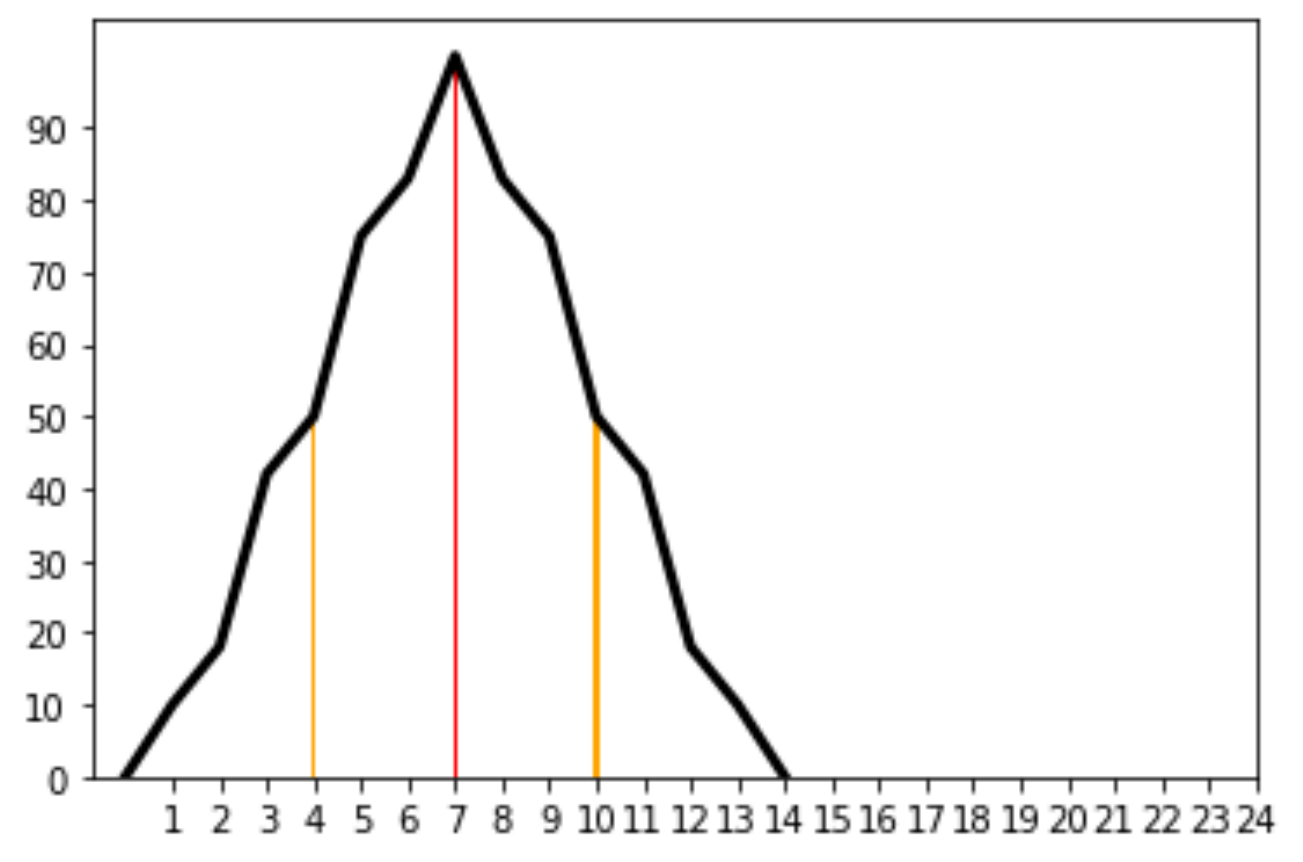}}
\resizebox*{4.4 cm}{!}{\includegraphics{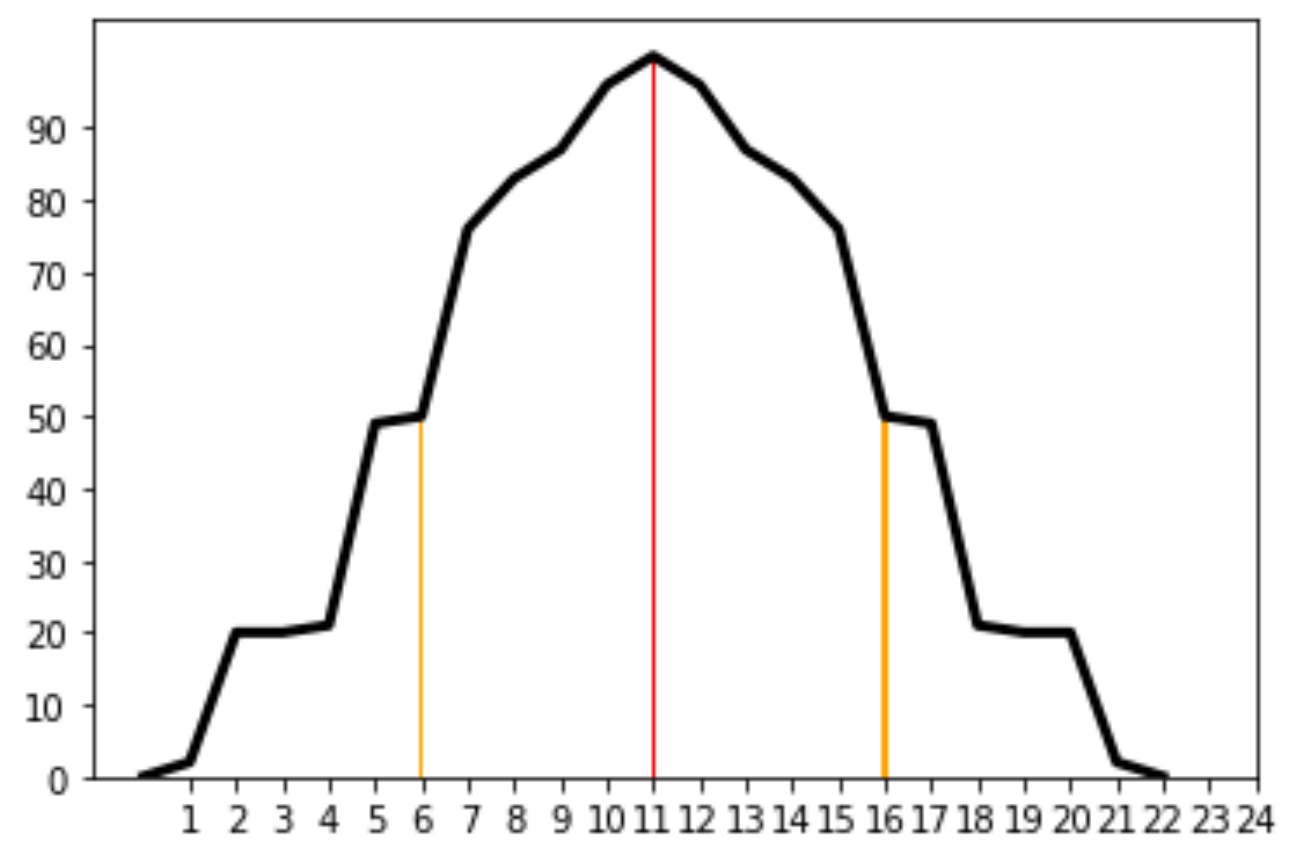}}
\caption{Different curves that describe the evolution of the viral load inside different hosts after the infection.} \label{viralload}
\end{figure*}

\begin{table*}[t!]
\centering
\renewcommand{\arraystretch}{1.6}
\caption{Probabilities of actions and corresponding percentage of emitted droplets across the evolution of the viral load $vl$.}
{\begin{tabular}{|c|c|c|c|c|c|}\hline

\textbf{Percentage of Viral Load ($vl$)} & \textbf{Breathing} & \textbf{Coughing} & \textbf{Sneezing} & \textbf{Talking} & \textbf{Perc(ID)} \\ \hline
\textcolor{blue}{$vl=0$}                          & 0.8                      & 0.1                     & 0.05                    & 0.05                                                                 & 0.1                                      \\ \hline
\textcolor{orange}{$0 < vl \leq 10$}            & 0.8                      & 0.1                     & 0.05                    & 0.05                                                                 & 0.15                                     \\ \hline
\textcolor{orange}{$10 < vl \leq 20$}             & 0.8                      & 0.1                     & 0.05                    & 0.05                                                                 & 0.2                                      \\ \hline
\textcolor{orange}{$20 < vl \leq 30$}             & 0.8                      & 0.1                     & 0.05                    & 0.05                                                                 & 0.2                                      \\ \hline
\textcolor{orange}{$30 < vl \leq 40$}              & 0.8                      & 0.1                     & 0.05                    & 0.05                                                                 & 0.2                                      \\ \hline
\textcolor{red}{$40 < vl \leq 50$}              & 0.75                     & 0.15                    & 0.05                    & 0.05                                                                 & 0.3                                      \\ \hline
\textcolor{red}{$50 < vl \leq 60$}              & 0.70                     & 0.20                    & 0.05                    & 0.05                                                                 & 0.4                                      \\ \hline
\textcolor{red}{$60 < vl \leq 70$}              & 0.65                     & 0.25                    & 0.05                    & 0.05                                                                 & 0.5                                      \\ \hline
\textcolor{red}{$70 < vl \leq 80$}             & 0.60                     & 0.30                    & 0.05                    & 0.05                                                                 & 0.6                                      \\ \hline
\textcolor{red}{$80 < vl \leq 90$}             & 0.55                     & 0.35                    & 0.05                    & 0.05                                                                 & 0.7                                      \\ \hline
\textcolor{red}{$\phantom{1}90 < vl \leq 100$}             & 0.50                     & 0.40                    & 0.05                    & 0.05                                                                 & 0.8                                     \\ \hline                                                
\end{tabular}}
\label{prob-table-b}
\end{table*}

\subsection{Propagation Model}
\label{Propagation Model}
Next, we present the third model of our framework, i.e., the propagation mode, discussing the underlying epidemic properties that describe the potential states an individual may be during the spread epidemic, possible actions of each individual that affect the transmission of an airborne pathogen, and the evolution of  the viral load inside the organism of an infected host.

\subsubsection{\textbf{Underlying Epidemic Model}}

The proposed approach for the modeling of the spread of an airborne pathogen, in our case the SARS-CoV-2, has its structure based on the characteristic behavior exhibited by the pathogen considering its transmission. In our approach we consider the SIII epidemic model with the following states:

\begin{enumerate}[i]
\item {\tt Susceptible}: individuals have never hosted the virus, being susceptible to it,
\item {\tt Infected}: individuals that got infected by the virus but are not yet contagious,
\item {\tt Infectious}: individuals who are infected and contagious, and 
\item {\tt Immune}: individuals that have recovered, and being neither Susceptible nor Infectious, they do not transmit the virus. 
\end{enumerate}
	 
Investigating the states that are critical for the contagion of the virus, in this work we study four possible action that can be performed by any individual in order to transmit an airborne pathogen. The possible actions considered by our model namely are {\tt breathing}, {\tt coughing}, {\tt sneezing} and {\tt talking} and each one, has a specific probability of being performed. At any time, the corresponding percentage of the viral load to be propagate through the infected emitted droplets is analogous to the evolution of the viral load inside the infected host. The probability of each action differs according to the epidemic state of each individual, as also to the corresponding viral load and the characteristics of the pathogen under consideration, in our case the SARS-CoV-2 virus. 

For example, an infectious individual exhibits an increased probability of coughing compared to an individual whose state is susceptible. In addition, the percentage of the viral load contained in the droplets emitted through an action performed by an individual is also relative to the epidemic state and the corresponding viral load. Moreover, the probability of making an action and the amount of the viral load transmitted by the droplets emitted through an action among individuals of the same epidemic state varies according to the evolution rate of the viral load inside different hosts.

In Tables~\ref{prob-table-a} and ~\ref{prob-table-b}, we provide an estimation of the probability of an individual to perform and action that transmits a portion of infected droplets that carry the airborne pathogen, in our case the SARS-CoV-2, and additionally we present an estimation on how this probability evolves across the evolution of the viral load inside an infected host, and the corresponding percentage of infected droplets. 

In particular, in Table~\ref{prob-table-a} we present the probability of an individual to perform one of the set of basic actions that mostly favor the transmission of an airborne pathogen through the emitted droplets, namely breathing, coughing, sneezing or talking, alongside with the corresponding volume of emitted droplets per action $VoED$. Moreover, in Table~\ref{prob-table-b} we present the evolution of the viral load ($vl$) inside an infected individual represented with values in the range $[0,100]$  (see, column 1), the  corresponding probabilities of performing an action that emits droplets (see, columns 2-4), and finally, concerning the evolution of the viral load inside the host individual, the corresponding percentage of infected droplets $Perc(ID)$ emitted, that are independent of the performed action (see, column 6).

\vspace{0.1 in}
\subsubsection{\textbf{Evolution of the Viral Load}}
Next, we discuss how the immune system in each individual behaves in general while hosting the airborne pathogen from the start of infection and how the corresponding viral load evolution varies inside different hosts and hence different immune systems, as it is of major importance to contain the aspect that the same pathogen may cause different results to different individuals. For example, the viral load evolution in an individual with chronical deceases is more aggressive than a young and healthy person who got infected. 

To include this case in our model we assign to each individual a behavioral profile, representing the viral load evolution by the time of infection. The corresponding viral load evolution curve that depicts this behavioral profile represents the information regarding the increasing ratio of the virus into the corresponding host, the first day of the symptoms and the viral load distribution during the days of being in infectious state. The constructed curves stochastically vary, and they have been developed focusing to simulate different behaviors of the viral load evolution through different individuals. At the end of this procedure we achieve the different confrontation of the virus on each individual maintaining the base features of the corresponding virus. 

As long as a susceptible individual gets infected, the stochastically developed behavioral profile describes the viral load evolution during the days, determining the incubation period, the first day of symptoms, the day that the viral load reaches the peak and the cure day (see, Figure~\ref{viralload}, where $x$-axis refers to the days after the day of infection and the $y$-axis refers to the percentage of the viral load). Taking into account the results presented in \cite{ref46}, we stochastically simulate the curve of viral load evolution estimating that a host is about to reach a $50\%$ of the viral load exhibiting the first symptoms $3$-$6$ days after the infection. From the day of symptoms onset, there is a margin of $3$-$6$ days for the viral load to increase to its peak. 

In our model, we consider that the decrease ratio of the viral load curve equals the corresponding increase ratio. As long as the percentage of the viral load of the corresponding infectious individual increases, the probability for coughing and the percentage of the infected respiratory droplets increases analogously. As a result, the individual becomes even more contagious and the probability to transmit the airborne pathogen increases too.

\begin{figure*}[t!]
\begin{center}
\resizebox*{5 cm}{!}{\includegraphics{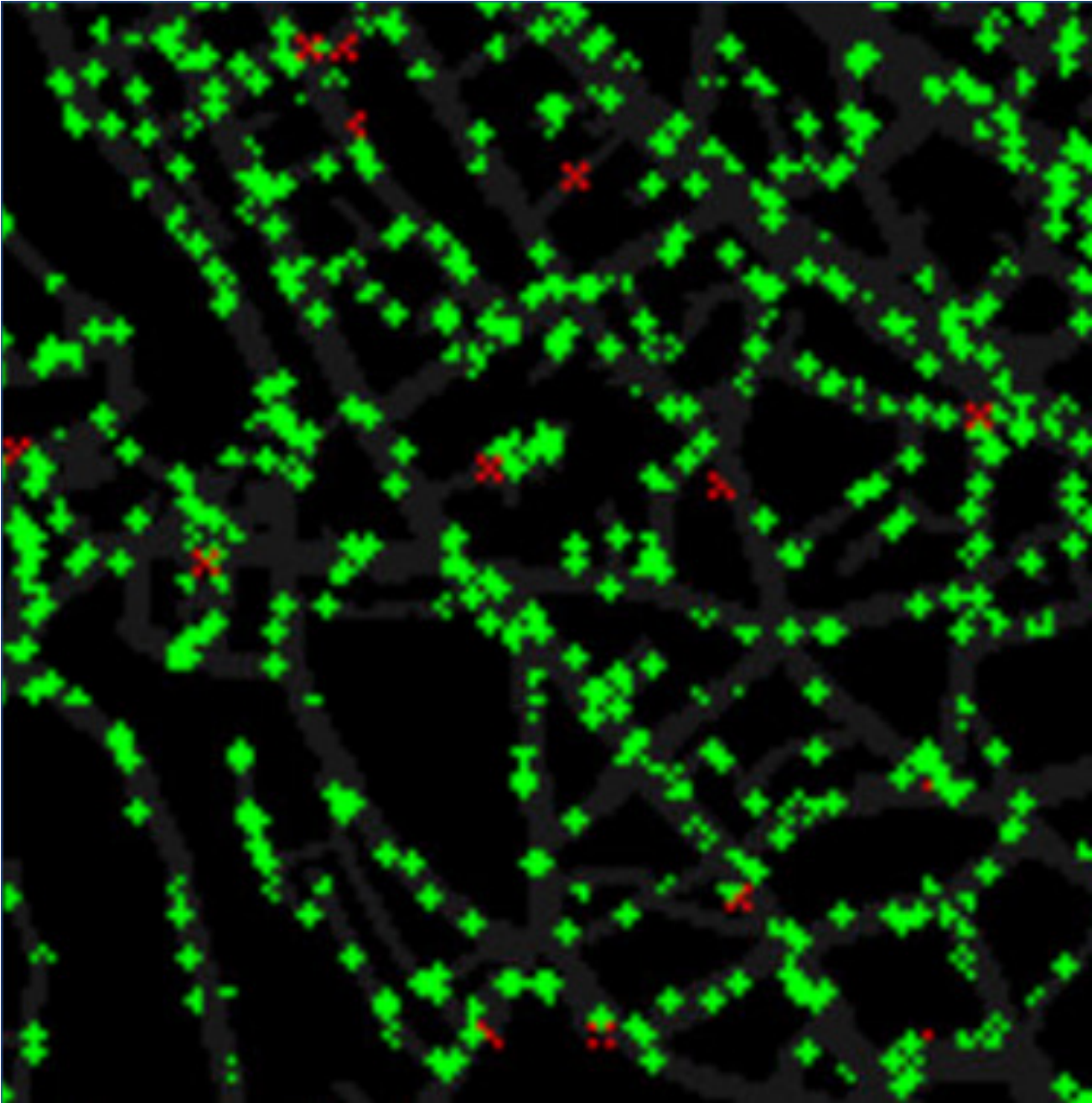}}\hspace{0.2 in}
\resizebox*{5 cm}{!}{\includegraphics{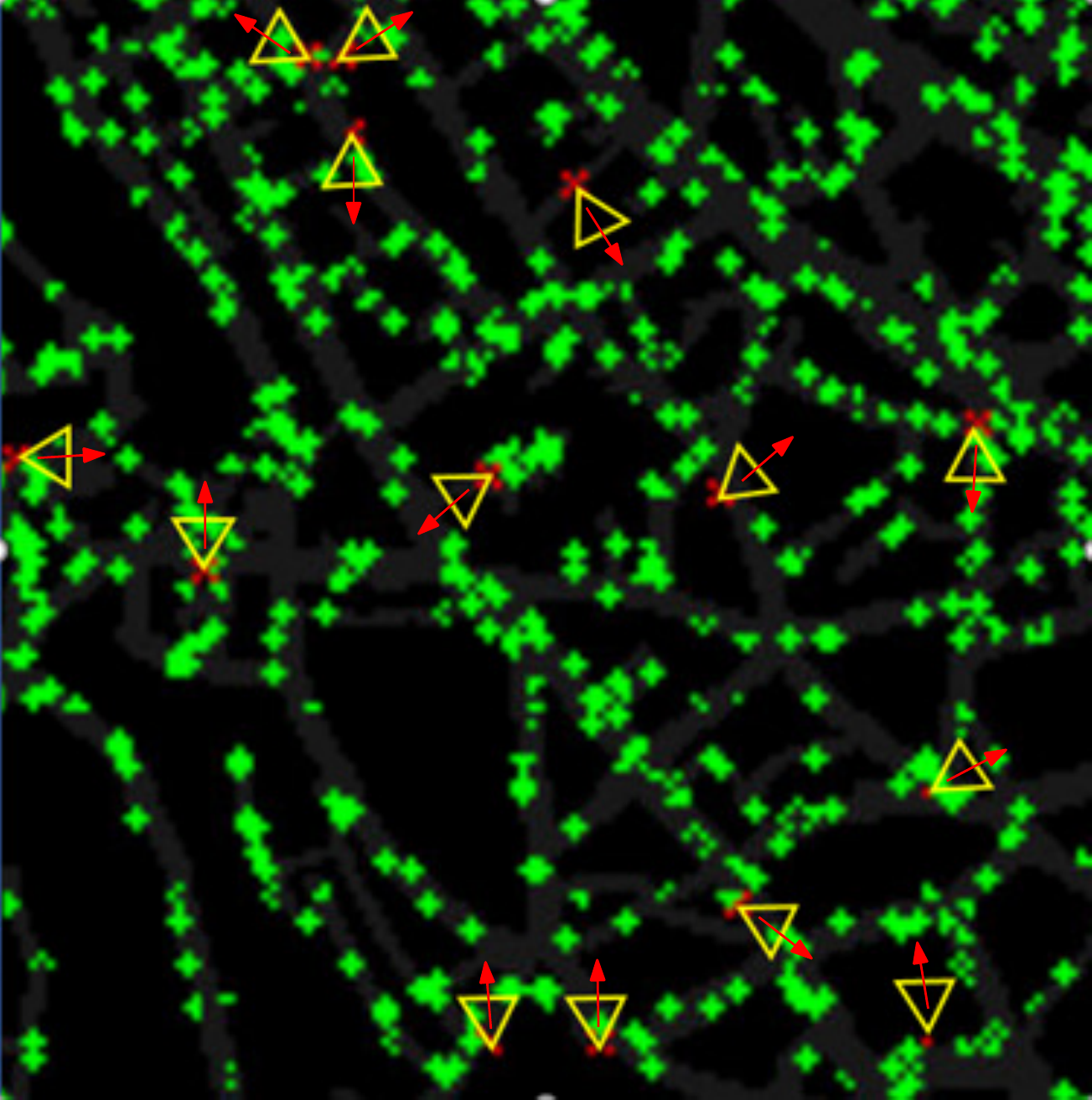}}\hspace{0.2 in}
\resizebox*{5.05 cm}{!}{\includegraphics{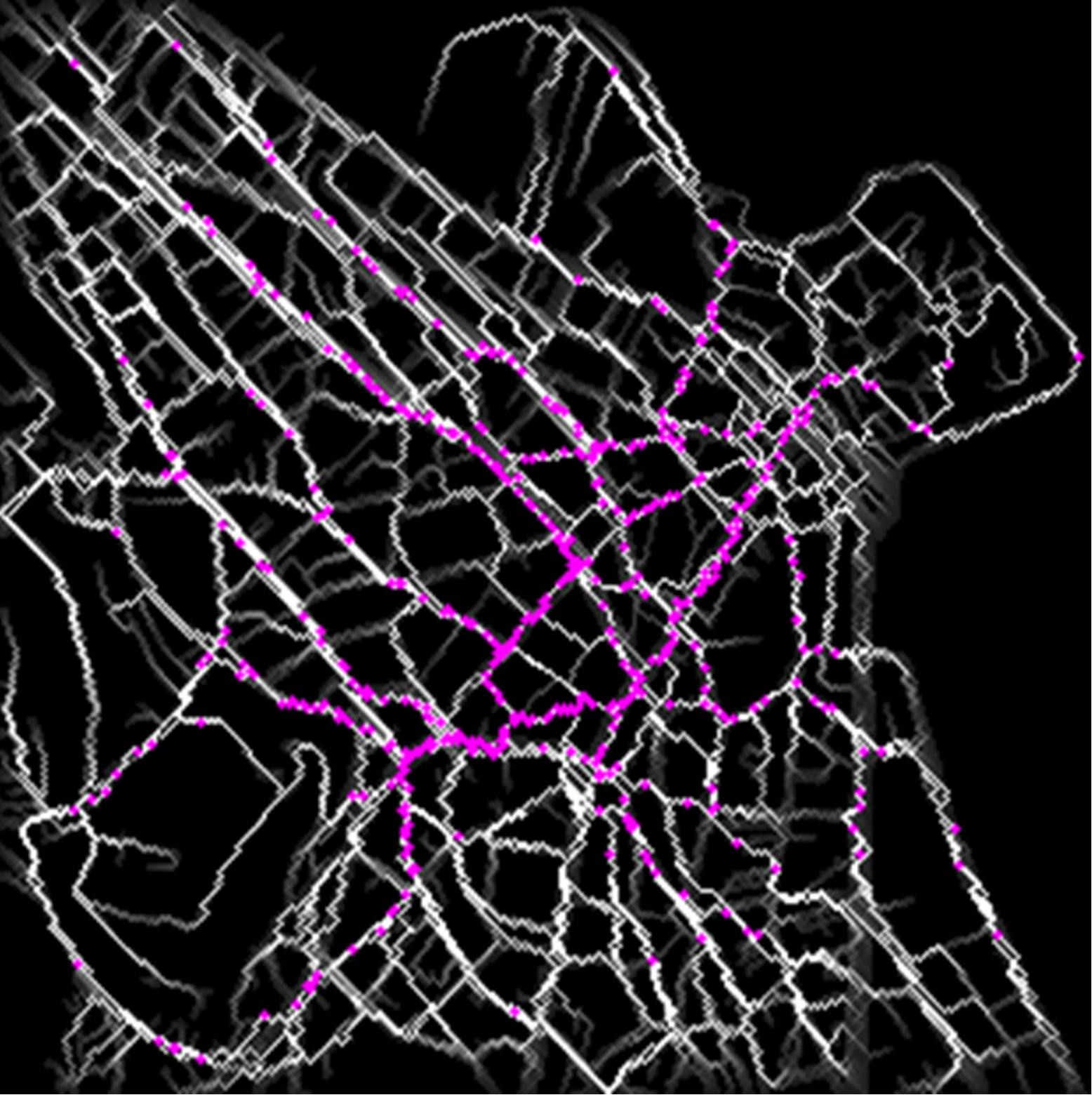}}\hspace{0.2 in}
\end{center}
\phantom{xxxxxxx}(a) Distribution of Individuals. \phantom{xxxxx}(b) Triangular cloud transmission. \phantom{xxxxxx}(c) Recorded Infections.
\caption{Airborne virus transmission from nearby infectious individuals in a specific area of the city, and points of registered potential infections.} \label{distrotriangles}
\end{figure*}

\subsubsection{\textbf{Infection of Proximal Population}}

Modeling the aspect of infection of proximal population, we focus into the physical substance of the procedure that describes the transmission of an airborne pathogen from an infectious individual to the nearby susceptible ones. Airborne pathogens spread through the air via respiratory droplets produced after an action performed by an infected individual (i.e., talk, breathe, cough, sneeze). The quantity of the produced droplets is proportional to the action performed, while the radius of each droplet can vary, depending on the intensity of the corresponding action, i.e., the produced droplets after a sneeze are up to $40000$, while coughing may produce up to $3000$ \cite{ref_drop3,ref_drop1,ref_drop2}.

Investigating the insight of this procedure, the proposed model considers the increased spread of the droplets over distance as the droplets transition in the air can be described as a triangular cloud whose one corner located in front of the infected individual, where the droplets released. To make our model more realistic, we consider the orientation of each individual into the $G_{map}$ graph developing over each individual a triangular cloud inside the boundaries of which the pathogen can be transmitted through the droplets emitted through an action performed by the infected individual. 
 
To simulate the transmission of the airborne pathogen and the infection of proximal population we need to consider the distance among them, the viral load and the type of the action performed (i.e., breath, talk, cough, sneeze). We develop a formula to calculate the probability of infection incorporating those three factors. The probability of an individual $X$ to get infected $P(X\rightarrow Infected)$ decreases as long as distance from nearby infected individual $Y$ increases, while respectively increases as long as viral load and the quantity of droplets produced are increased. 

The formula to compute the probability of an individual $X$ to get infected by the action of a nearby individual $Y$ that has a specific viral load and performs a specific action is computed as follows:

\begin{equation}\label{probinfeq}
P(X \rightarrow Infected)=\frac{Perc(ID)\times Perc(VoED)}{1+Eucl(X,Y)^2},                          
\end{equation}

where $Perc(ID)$ corresponds to the percentage of infected droplets a nearby infected individual emits by an action, $Perc(VoED$) corresponds to the volume $V$ of the emitted droplets a nearby infected individual emits by an action, and the $Eucl(X,Y)^2$ corresponds to the square of their in-between Euclidean Distance.

\begin{figure}[t!]
\centering
\includegraphics[scale=0.38]{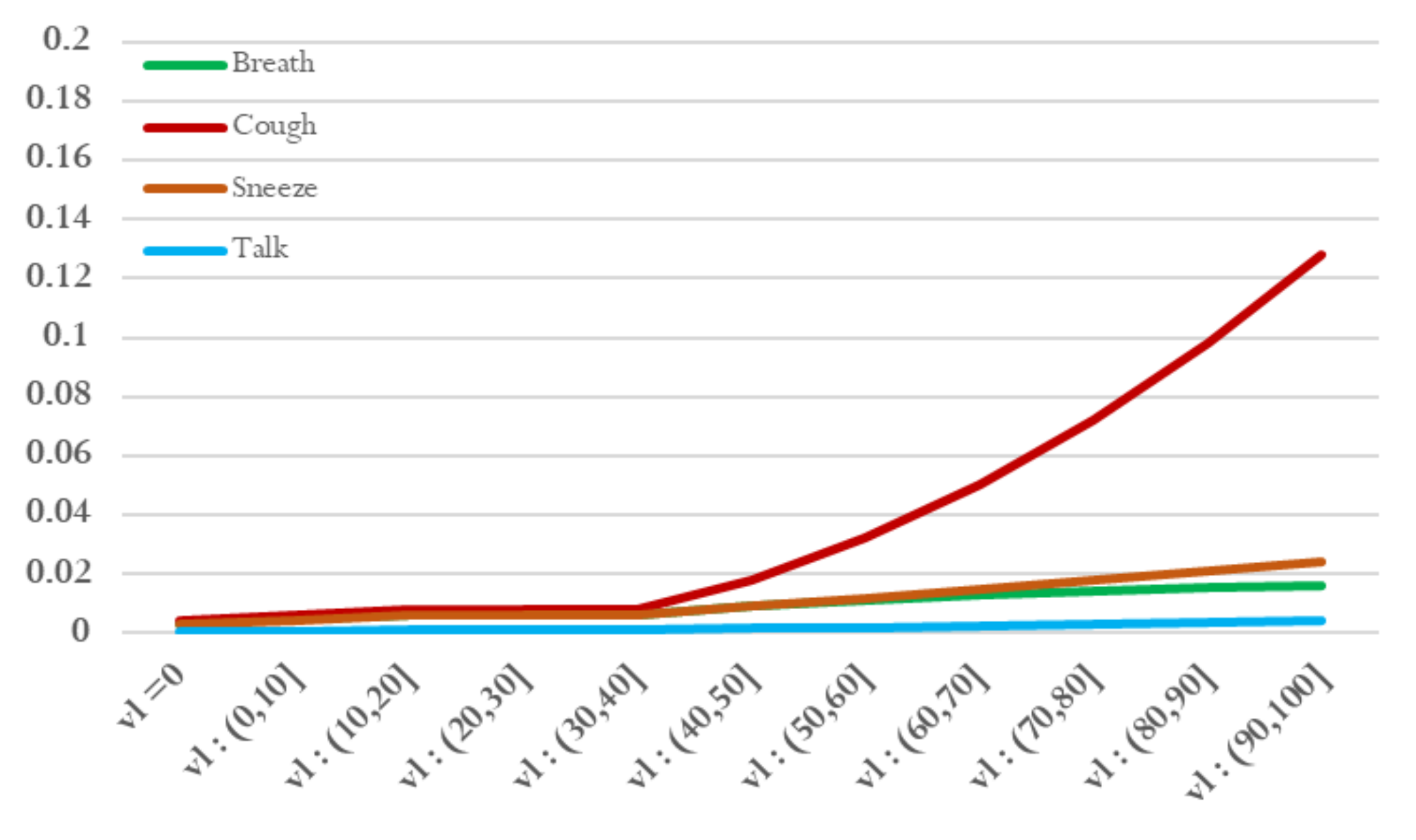}
\caption{Probability of infection by a nearby host at distance $2m$ according to Table~\ref{prob-table-a}, Table~\ref{prob-table-b}, and Equation\ref{probinfeq}} \label{probinf}
\end{figure}

In Figure~\ref{distrotriangles} we illustrate an example of the utilization of the proposed model. Particularly, in Figure~\ref{distrotriangles}~(a) there are depicted the positions of the individuals inside the city area, where the green marks correspond to the susceptible individuals, while the red marks correspond to the infectious individuals. In Figure~\ref{distrotriangles}~(b) there are depicted the triangular cloud that contain the respiratory droplets emitted through the actions performed by the infected individuals towards their direction according to their moves, while in Figure~\ref{distrotriangles}~(c) we represent with pink marks the points of the area where the infections were recorded according to our simulations experiments.

At this point, it is of major importance to note that the accumulation of the points where infections were recorded exhibits an identical density exactly over the most frequently walked street points (see, Figure~\ref{routes2}), which is a fact expected from our reality-based intuition, attesting thus the precision of our proposed model. 

Finally, in Figure~\ref{probinf} we combine the information from Table~\ref{prob-table-a}, Table~\ref{prob-table-b}, and utilizing Equation~\ref{probinfeq} we represent the probability of infection by a nearby infected individual that has a safety distance of $2m$ \cite{distance} from a susceptible individual. For example, let as assume two infectious individuals with the same amount of viral load and same orientation, let as say 5 meters from two susceptible individuals respectively. The first infectious individual sneezes while the second one coughs. The probability that the first infectious individual will transmit the virus to the corresponding susceptible individuals is higher compared to the second one, considering the more quantity of respiratory droplets produced by sneezing over coughing while the distance, orientation and viral load remains the same.

\section{Conclusion}
\label{Conclusion}
In this work we  proposed and implemented a stochastic graph-based model for the simulation of SARS-Cov-2 spread across individuals that move into a specific area. The proposed design incorporated three sub-models, namely the spatial, the mobility, and the propagation models, in order to be incorporated for the integration of the simulation framework. We first presented the spatial model, that is utilized to construct the city basis and the points of interest from a Google Maps images. Additionally, we presented the mobility model which describes the schedule of the mobility of the individuals, their shortest path calculation from a source to a destination point using shortest path algorithms and their multiple tasks during their navigation through the initial city. Moreover, we presented the propagation model which describes the underlying epidemic model, the viral load evolution inside a host and the infection over the nearby susceptible population considering the orientation of the corresponding infected individual, with respect to the properties of the airborne virus SARS-CoV-2. 

Finally, concerning the future research, our main goals are pointed to the investigation of several factors that affect the spread of an airborne pathogen regarding the application of countermeasures, that could be potentially deployed in order to reduce the transmission of a pathogen among the susceptible population. The deployed countermeasures that are about to be investigated, range between the application of ``density reduction", ``self-quarantine", ``contact tracing", and combinations of them, as also to the augmentation of the contact tracing procedures through the utilization of efficient contact tracing algorithms, e.g. the $2D-$Co-Contact Algorithm proposed in \cite{refTR}, in order to provide evidences about the potentials of our model when deployed for epidemic control and pandemic prevention. Throughout this aspect we estimate that the proposed approach could be also utilized in order to provide information about the dynamics of an epidemic caused by an airborne pathogen consisting hence a valuable tool against the mitigation of future pandemics.


\end{document}